\documentclass[aps,fullpaper,twocolumn,showpacs,amsmath,amssymb,groupedaddress]{revtex4-1}

\usepackage{graphicx}
\usepackage{dcolumn}
\usepackage{bm}
\usepackage{color}
\usepackage{ulem}

\usepackage{dcolumn}
\usepackage{bm}

\usepackage{here}

\begin{document}

\title{
Spontaneous formation of kagome network and Dirac half-semimetal on a triangular lattice 
}

\author{Yutaka Akagi$^{1,2}$\thanks{E-mail address: yutaka.akagi@oist.jp} and Yukitoshi Motome$^2$}

\affiliation{
$^1$Okinawa Institute of Science and Technology, Onna-son, Okinawa 904-0412, Japan\\
$^2$Department of Applied Physics, University of Tokyo, Tokyo 113-8656, Japan}

\date{\today}

\begin{abstract}
In spin-charge coupled systems, geometrical frustration of underlying lattice structures can 
give rise to nontrivial magnetic orders and electronic states.
Here we explore such a possibility in the Kondo lattice model with classical localized spins on a triangular lattice by using a variational calculation and simulated annealing. 
We find that the system exhibits a four-sublattice collinear ferrimagnetic phase at 5/8 filling for a large Hund's-rule coupling. 
In this state, the system spontaneously differentiates into the up-spin kagome network and the isolated down-spin sites, which we call the kagome network formation. 
In the kagome network state, the system becomes Dirac half-semimetallic: 
The electronic structure shows a massless Dirac
node at the Fermi level, and the Dirac electrons are almost fully spin polarized due to the large Hund's-rule coupling.
We also study the effect of off-site Coulomb repulsion in the kagome network phase where the system is effectively regarded as a 1/3-filling spinless fermion system on the kagome lattice. 
We find that, at the level of the mean-field approximation, a $\sqrt{3} \times \sqrt{3}$-type 
charge order occurs in the kagome network state, 
implying the possibility of fractional charge excitations in this triangular lattice system. 
Moreover, we demonstrate that the kagome network formation with fully-polarized Dirac electrons are controllable by an external magnetic field. 
\end{abstract}


\maketitle

\section{ \label{sec:intro}
Introduction
}

The ferromagnetic (FM) Kondo lattice model, often called the double-exchange (DE) model~\cite{Zener1951,Anderson1955,de-Gennes_1960}, 
is one of the fundamental models for correlated electron systems. 
It has been extensively studied for a long time, 
mainly for understanding of the physical properties of perovskite manganese oxides~\cite{Kaplan1999,Tokura1999}. 
The phase transition to the FM metallic state by decreasing temperature 
and the colossal negative magnetoresistance are well explained by the DE mechanism: 
An effective FM interaction is induced by the kinetic motion of electrons 
under the influence of the large Hund's-rule coupling to localized magnetic moments.
On the other hand, the competition between the FM DE interaction 
and the antiferromagnetic (AFM) superexchange (SE) interaction between the localized spins 
has also been studied intensively. 
In the early stage, a spin canting state with a spin-flop type ordering was predicted 
in the lightly doped region, which smoothly connects the AFM insulating state at half filling  
and a hole-doped FM metal~\cite{de-Gennes_1960}. 
Later, the scenario was revisited; 
a phase separation (PS) was found to take place between the AFM insulator and FM metal, 
and hinders the canting state~\cite{Yunoki_1998,Dagotto_2001}. 
In addition, at the commensurate 1/4 filling, 
the magnetic competition leads to the first-order transition 
from the FM metal to an insulator with a peculiar ``flux"-type magnetic ordering~\cite{Yamanaka1998,Agterberg2000}.

Recently, the FM Kondo lattice model has attracted renewed interest 
from the viewpoint of geometrical frustration in underlying lattice structures. 
In general, the geometrical frustration leads to competition between different magnetic orders, 
resulting in peculiar states, such as a complicated ordering, liquid-like, and glassy states. 
Such peculiar magnetism significantly affects the electronic state of itinerant electrons 
and gives rise to peculiar transport phenomena. 
A typical example is the unconventional anomalous (or topological) Hall effect 
discovered in some pyrochlore oxides~\cite{Taguchi2001,Nakatsuji2006,Machida2007}. 
Theoretically, the topological Hall effect is caused by the coupling of itinerant electrons 
to a noncoplanar spin configuration with 
nonzero spin scalar chirality~\cite{Loss1992,Ye1999,Tatara2002}. 
Indeed, such a mechanism was investigated in many frustrated systems, 
such as triangular~\cite{Martin2008,Akagi2010,Kato2010}, 
kagome~\cite{Ohgushi2000,Mathieu2006,Ishizuka2013a,Rahmani2013, Barros2014}, 
face-centered-cubic~\cite{Shindou2001}, 
and pyrochlore lattices~\cite{Udagawa2013}. 
Fermi surface properties are important for stabilizing such noncoplanar orders~\cite{Akagi2012,Hayami2014_FS}. 
Other interesting phenomena are resistivity minimum~{\cite{Nakatsuji2006,Sakata2011} 
and bad-metallic behavior~\cite{Iguchi2009,Iguchi2011}. 
These peculiar transport properties were also discussed theoretically~\cite{Udagawa2012,Chern2013,Motome2010a,Kumar2010}.

The competition between the FM DE interaction and the AFM SE interaction 
has also been studied in geometrically frustrated systems. 
For instance, a variety of magnetic phases were predicted by the mean-field analysis 
in kagome and pyrochlore lattice systems~\cite{Ikoma2003}. 
Monte Carlo studies were done for a pyrochlore lattice system, 
unveiled bad-metallic behavior~\cite{Motome2010a}, 
a peculiar PS~\cite{Motome2010b}, and a spontaneous 
spin Hall effect by inversion symmetry breaking~\cite{Ishizuka2013b}. 
Triangular and checkerboard lattice systems were also studied by Monte Carlo simulation, 
and scalar chiral ordered phases were found~\cite{Kumar2010,Venderbos2012}.
Furthermore, the low density region
in the triangular lattice system was studied by 
variational calculations, 
and a noncoplanar three-sublattice spin canting order was found 
adjacent to a PS~\cite{Akagi2011}. 
These results indicate that the competition between the FM DE and AFM SE interactions 
leads to much richer physics in geometrically frustrated systems.

In this paper, we investigate magnetic and electronic phases induced by 
the competition between the FM DE 
and AFM SE interactions on a triangular lattice.
We study the ground state of the FM Kondo lattice model with the AFM SE interaction 
by complementarily using the variational calculation and the simulated annealing. 
We find that a four-sublattice collinear ferrimagnetic phase appears 
in the large Hund's-rule coupling region at a commensurate 5/8 filling. 
In this phase, the triangular lattice system is spontaneously differentiated 
into a kagome network (KN) composed of up-spin sites and isolated down-spin sites. 
Under the KN formation, the electronic structure shows a Dirac node with linear dispersion, and the Fermi level at 5/8 filling is located just at the Dirac node. 
The Dirac electrons are almost fully spin polarized by the large Hund's-rule coupling. 
A similar Dirac half-semimetal was also found at 1/3 filling in a different ferrimagnetic phase 
in the absence of the AFM SE interaction~\cite{Ishizuka2012}. 
In our KN state, however, excess electrons over half filling are effectively regarded as spinless fermions on the kagome lattice at 1/3 filling. 
This leads us to study the effect of the off-site Coulomb repulsive interaction between itinerant electrons 
in the KN phase from the interest of a fractional charge excitation on a triangular lattice. 
We also clarify the effect of an external magnetic field on the KN formation with half-semimetallic Dirac electrons. 

The organization of this paper is as follows.
In Sec.~\ref{sec:model_and_method}, we introduce the model and theoretical methods used in this paper.
After introducing the FM Kondo lattice model in Sec.~\ref{sec:madel}, 
we describe the details of the variational calculation and simulated annealing in Secs.~\ref{sec:variational_calculation_method} and \ref{sec:simulated_annealing}, respectively. 
In Sec.~\ref{sec:result}, we present the results for the KN phase. 
We discuss the ground-state phase diagram obtained by the variational calculation (Sec.~\ref{sec:vatiational_calculation}), the electronic structure in the KN phase (Sec.~\ref{sec:Dirac}), and the stability of the KN phase examined by the simulated annealing (Sec.~\ref{sec:annealing_result}). 
In Sec.~\ref{sec:mean-field_result}, we present the mean-field results for the effect of the electron-electron interaction between nearest-neighbor sites 
and discuss the possibility of fractional charge excitations.
Moreover, we discuss a controllability of
the Dirac half-semimetal, magnetic supersolid, and 
fractional charge excitations by an external magnetic field in Sec.~\ref{sec:magnetic-field}. 
Finally, Sec.~\ref{sec:summary} is devoted to a summary.
In the Appendix~\ref{app:PD_single-ion-anisotropy}, we present the results of variational calculations for the effect of single-ion anisotropy on the KN phase.

\section{ \label{sec:model_and_method}
Model and Method
}
In this section, we introduce the model and methods. The model is given in Sec.~\ref{sec:madel}. 
Details of the variational calculation and the simulated annealing are described in 
Secs.~\ref{sec:variational_calculation_method} and \ref{sec:simulated_annealing},
respectively.

\subsection{ \label{sec:madel}
Kondo lattice model
}

We consider a Kondo lattice model on a triangular lattice, whose Hamiltonian is given by
\begin{align}
{\cal H}_{\rm KL}=&-t \sum_{\langle i,j \rangle,\alpha } ( c^{\dagger}_{i,\alpha } c_{j,\alpha }+\mathrm{h.c.}) \notag\\ 
&-J_{\rm H} \sum_{i,\alpha ,\beta } 
    c^{\dagger}_{i,\alpha } \boldsymbol{\sigma}_{\alpha \beta }  c_{i,\beta } \cdot \mathbf{S}_i
+J_{\rm AF} \sum_{\langle i,j\rangle} \mathbf{S}_i \cdot \mathbf{S}_j. 
\label{Ham:DE}
\end{align}
The first term describes the hopping of itinerant electrons, 
$c^{\dagger}_{i,\alpha }$($c_{i,\alpha }$) is a creation (annihilation) operator for an itinerant 
electron with spin $\alpha$ on site \textit{i},
and $t$ is the transfer integral. 
The sum $\langle i,j\rangle$ is taken over the nearest-neighbor sites on the triangular lattice.
The second term denotes the FM Hund's-rule coupling between itinerant electron spins and localized spins,
$J_{\rm H}$ is the coupling constant, 
$\boldsymbol{\sigma}_{\alpha \beta }=({\sigma}^x_{\alpha \beta },{\sigma}^y_{\alpha \beta },{\sigma}^z_{\alpha \beta })$
are Pauli matrices, and ${\mathbf S}_i$ is a localized spin on site \textit{i}. 
The last term describes the AFM SE interaction with the coupling constant $J_{\rm AF}$ between the nearest-neighbor localized spins ($J_{\rm AF} \ge 0$).
Here, we consider classical spins for ${\mathbf S}_i$ with $|{\mathbf S}_i|=1$.
Note that the sign of $J_{\rm H}$ is irrelevant 
unless localized spins are quantum ones.
Hereafter, we take $t=1$ as the unit of energy, the lattice constant $a=1$, and the Boltzmann constant $k_{\rm B}=1$.

\begin{figure*}[!htbp]
\begin{center}
\includegraphics[width=1.95\columnwidth]{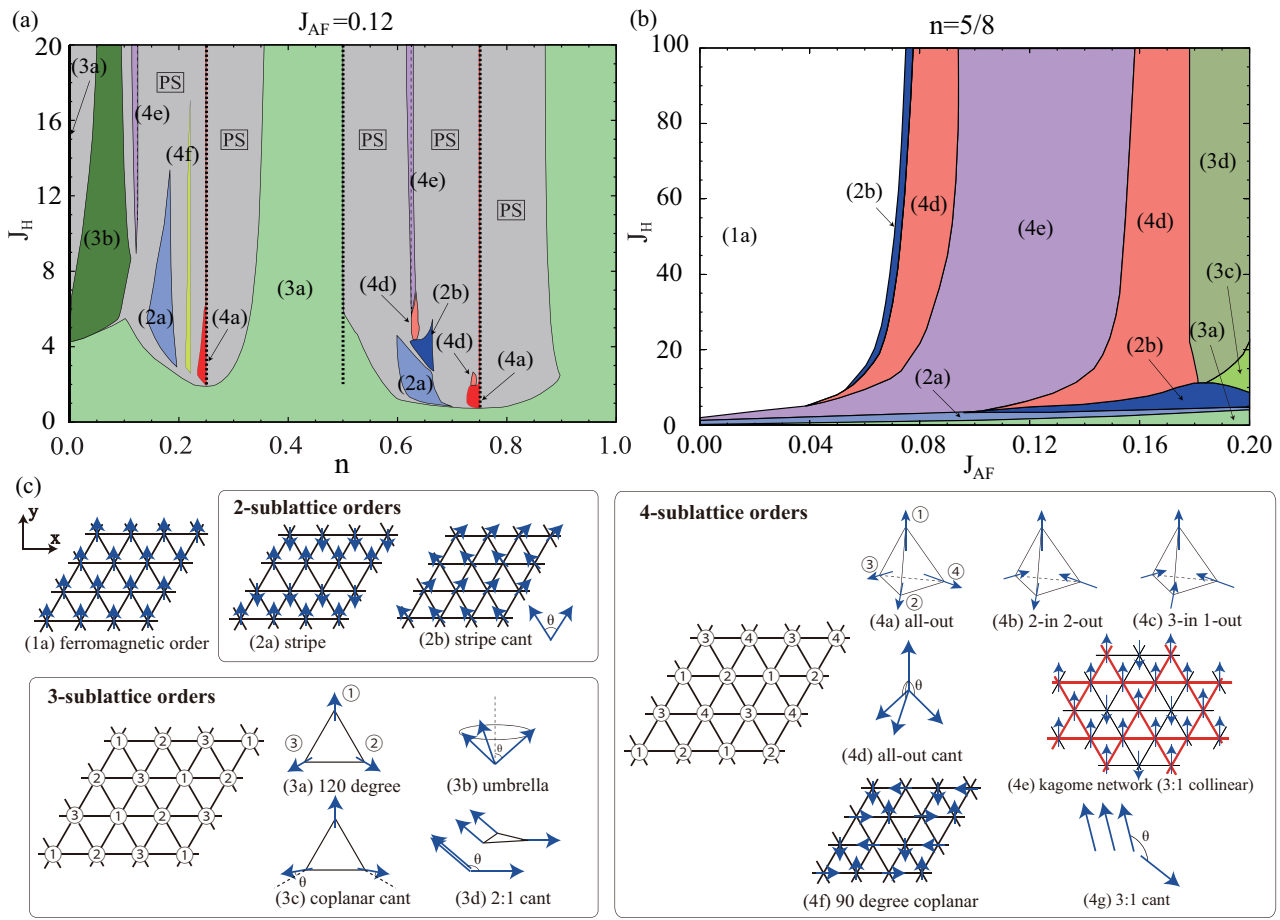}
\end{center}
\caption{(Color online) Ground-state phase diagrams for the model (\ref{Ham:DE}) on a triangular lattice as functions of (a) $n$ and $J_{\rm{H}}$ at $J_{\rm{AF}}=0.12$, and
(b) $J_{\rm{AF}}$ and $J_{\rm{H}}$ at $n=5/8$. 
The vertical thick (thin) dashed lines at $n=1/4$, $1/2$, and $3/4$ ($n=1/8$ and $5/8$) in (a) indicate gapful insulating 
(zero-gap semiconducting)
regions. PS stands for the phase-separated region
(see the text in Sec.~\ref{sec:variational_calculation_method} for details). 
The thick black line at $n = 0$ represents that the (3a) $120^{\circ }$ coplanar ordered phase is stabilized at zero doping.
In (b), we ignore the possibility of PS. 
(c) Ordering patterns of localized spins used in the variational calculations: 
(1a) FM order, (2a) and (2b) two-sublattice orders, (3a)-(3d) three-sublattice orders, and (4a)-(4g) four-sublattice orders
(see the text in Sec.~\ref{sec:variational_calculation_method} for details).
The numbers in (3a) and (4a) indicate the sublattices at which the spins are located on the two-dimensional lattice shown
in each left.
}
\label{PD-variational-calc_all-order}
\end{figure*}

\subsection{ \label{sec:variational_calculation_method}
Variational calculation
}

In the present study, we 
are interested in the ground-state (zero-temperature) phase diagram of the model in Eq.~(\ref{Ham:DE}) in the parameter space of the electron filling 
$n=\frac{1}{2N}\sum_{i\alpha }\langle c^{\dagger}_{i,\alpha } c_{i,\alpha }\rangle$ 
(\textit{N} is the total number of sites), $J_{\rm H}$, and $J_{\rm AF}$ 
[see Figs.~\ref{PD-variational-calc_all-order}(a) and \ref{PD-variational-calc_all-order}(b)]. 
In order to obtain the global structure of the phase diagram, 
we adopted a variational method in which 
we compare the ground state energies of possible ordered states.
The method is the same as those used in the previous studies~\cite{Akagi2010,Akagi2011}, but we describe the details below for making the paper self-contained. 
The method is just used for readily searching parameter regions 
where possible interesting phases emerge.
As a complementary method, we adopt the simulated annealing (Sec.~\ref{sec:simulated_annealing}) to check the stability of the remarkable 
state that we are particularly interested in, the KN state (4e) 
in Fig.~\ref{PD-variational-calc_all-order}(c).

For the current model on the triangular lattice,
we consider 14 different types of ordered states 
up to four-site unit cell, as shown in Fig.~\ref{PD-variational-calc_all-order}(c).
Figure~\ref{PD-variational-calc_all-order}(c)(1a) shows a FM order.
Figures~\ref{PD-variational-calc_all-order}(c)(2a) and
\ref{PD-variational-calc_all-order}(c)(2b) show two-sublattice orders: (2a) a collinear stripe order and (2b) a stripe order with a canting angle $\theta $.
Figures \ref{PD-variational-calc_all-order}(c)(3a)-\ref{PD-variational-calc_all-order}(c)(3d) show three-sublattice orders: (3a) a $120^{\circ }$ noncollinear order,
(3b) a noncoplanar umbrella-type order with angle $\theta $
(canted in the normal direction to the coplanar plane from the $120^{\circ }$ order) which was discussed in Ref.~\cite{Akagi2011},
(3c) a coplanar order with canting angle $\theta $ for two spins from $120^{\circ }$ order,
and (3d) a 2:1-type order with two parallel spins that have angle $\theta $ to the remaining one.
Figures \ref{PD-variational-calc_all-order}(c)(4a)-\ref{PD-variational-calc_all-order}(c)(4g) show four-sublattice orders:
(4a) an all-out-type order which was discussed in detail in Ref.~\cite{Akagi2010} and \cite{Akagi2012},
(4b) a two-in two-out-type order,
(4c) a three-in one-out-type order,
(4d) an all-out-type order with canting angle $\theta $ for three spins,
(4e) a 3:1 collinear order,
(4f) a coplanar order with a $90^{\circ }$ flux-type configuration, and 
(4g) a 3:1 canted order which is a four-sublattice version of the (3d) phase.

Among the various ordered states, 
we are particularly interested in the four-sublattice ferrimagnetic order (4e) in this paper.
The magnetic ordering structure is collinear with three-up one-down spin configuration per four sites; 
the up-spin sites comprise a kagome lattice in the triangular lattice, and the down-spin sites are 
at the centers of the six-site hexagons in the up-spin kagome network, as shown in Fig.~\ref{PD-variational-calc_all-order}(c)(4e). 
We call this state (4e) the kagome network (KN) phase hereafter.

In the calculation of the energy for each state at zero temperature, 
we computed the integral over the first Brillouin zone 
by approximating it by the sum over grid points of $1600 \times 1600$. 
We carefully checked the possibility of electronic 
PS by comparing the grand potential as a function of the chemical potential~\cite{Akagi2010}. 
In general, in the spin-charge coupled systems like the present model in Eq.~(\ref{Ham:DE}), 
PS takes place between different magnetic phases, as the electron density jumps at the transition~\cite{Yunoki_1998, Dagotto_2001}. 
Thus, two phases located on each side of PS coexist in the PS region, e.g., the (3a) and (4a) phases exist together 
in the PS region for $3/4 < n \lesssim 0.87$ at $J_{\rm H}=20$.
For the states (2b), (3b), (3c), (3d), 
(4d), and (4g), we optimized the canting angle $\theta $.
Note that
(2b) with $\theta =\pi $, (3b) with $\theta =\frac{\pi }{2}$, (3c) with $\theta =0$,
(4d) with $\theta = \cos^{-1}(-1/3)$, $\theta = \cos^{-1}(+1/3)$, 
$\theta = \pi$, and (4g) with $\theta = \pi$ 
are equivalent to (2a), (3a), (3a), (4a), (4c), 
(4e), and (4e), respectively.
Although an incommensurate order might take place for a general filling,
we consider only uniform $\mathbf{q}=\mathbf{0}$ orders with the magnetic unit cells listed above.

\subsection{ \label{sec:simulated_annealing}
Simulated annealing 
}

In order to check the stability of the KN state (4e) against other ordered states 
having larger unit cells, we used the simulated annealing method.
The simulated annealing is a technique for the optimization 
of a given function in the large parameter space~\cite{Kirkpatrick1983}.
It is often used to obtain
a candidate for the ground state in complicated systems.
In the present case, we considered 
an enlarged magnetic unit cell including 12 sites (see the inset of Fig.~\ref{SA_SI}), and optimized 
the localized spin configuration by the simulated annealing. 
Namely, temperature $T$ is decreased gradually, and at each $T$, the spin configuration is updated by the Monte Carlo sampling.
The cooling process is done in the geometrical way, 
i.e., $T_{k+1}=\alpha T_k$ 
($0<\alpha<1$), where $T_k$ is the temperature in the $k$-th step. 
For instance, we take $\alpha = 0.93$, the initial temperature $T_1=0.1$, and 
$k = 132$ for the final step of cooling in the calculations 
(the final temperature is $T_{132}\simeq 7.4 \times 10^{-6}$) 
in Secs.~\ref{sec:annealing_result} and \ref{sec:magnetic-field}.

In the Monte Carlo sampling at each $T$, 
we adopted the standard technique used for the model in Eq.~(\ref{Ham:DE})~\cite{Yunoki_1998}.
The partition function of the present system is given by
$Z={\rm Tr}_{ \{\mathbf{S}_i\} } {\rm Tr}_{\rm F} \exp[-\beta ({\cal H}-\mu {\hat N}_{\rm e})]$,
where ${\rm Tr}_{ \{\mathbf{S}_i\} }$ and ${\rm Tr}_{\rm F}$ are traces 
over classical localized spin degrees of freedom
and fermion degrees of freedom, respectively.
Here, $\beta =1/T$ is the inverse temperature, $\mu$ is the chemical potential, and ${\hat N}_{\rm e}$ is the total number operator for fermions. 
In the Monte Carlo simulation,
${\rm Tr}_{ \{\mathbf{S}_i\} }$ is calculated by using the Markov-chain Monte Carlo sampling. 
The updates are done by the single-spin flip on the basis of the standard Metropolis algorithm.
In order to obtain the Monte Carlo weight, the trace ${\rm Tr}_{\rm F}$ is calculated
by the exact diagonalization of the Hamiltonian for a given spin configuration. 
In the calculation, we took the summation over grid points of $10\times 10$ in the (folded) first Brillouin zone, which means that the effective system size is $12 \times 10 \times 10$. 
It is crucial to take such a large system size in the calculations for the KN phase; 
since the KN phase has a Dirac node at the Fermi level as discussed in Sec.~\ref{sec:Dirac}, 
sufficiently dense grid points in the momentum space are necessary to take into account the energy contribution from the low-energy linear dispersion.

\section{ \label{sec:result}
Results and Discussions
}
In this section, we present the results 
obtained by the variational calculation and the simulated annealing 
introduced in Secs.~\ref{sec:variational_calculation_method} and \ref{sec:simulated_annealing}, respectively.
We first present the phase diagrams obtained by the variational calculation 
in Sec.~\ref{sec:vatiational_calculation}.
We show that the system exhibits the KN phase at 5/8 filling. 
In Sec.~\ref{sec:Dirac}, we discuss the characteristic electronic structure of the KN phase, the massless Dirac node. 
Next, we discuss the stability of the KN phase by the simulated annealing 
in Sec.~\ref{sec:annealing_result}.
In Sec.~\ref{sec:mean-field_result}, we consider the effect of the electron-electron interaction between nearest-neighbor sites, $V$, at the level of mean-field approximation. 
We present the phase diagram as a function 
of $V$ and $J_{\rm AF}$, and discuss the robustness of the KN phase and the possibility of fractional charge excitations. 
Finally, we discuss the effect of 
an external magnetic field on the KN state by using the simulated annealing 
in Sec.~\ref{sec:magnetic-field}.
We show that the system exhibits a continuous transition between the 
KN state and a magnetic supersolid.

\subsection{ \label{sec:vatiational_calculation}
Phase diagram by the variational calculation
}

Figure~\ref{PD-variational-calc_all-order}(a) shows the result of the ground-state phase diagram as a function of the electron 
filling $n$ and the Hund's-rule coupling $J_{\rm{H}}$ at $J_{\rm{AF}}=0.12$.
As the AFM SE interaction $J_{\rm{AF}}$ favors the $120^{\circ }$ N{\'e}el order on the triangular lattice, 
the (3a) state appears in a broad parameter region as shown in the figure.
The phase diagram shares many aspects with the previous results in Refs.~\cite{Akagi2010} and \cite{Akagi2011}: 
for instance, (4a) all-out chiral phases at $n=1/4$ and $3/4$, and (3b) umbrella-type ordered phase in the low density region.

The new finding, however, is the KN state (4e) emerging at $n=1/8$ and $5/8$. 
As described in Sec.~\ref{sec:variational_calculation_method}, 
the magnetic structure consists of the up-spin kagome network in the triangle lattice and 
the down-spins at the remaining sites, as shown in Fig.~\ref{PD-variational-calc_all-order}(c)(4e).
To further confirm the stability of the KN phase within the variational calculation, 
we consider all the possible magnetic orders in a larger 12-site unit cell (see the inset of Fig.~\ref{SA_SI}) 
by replacing the localized spins by the Ising spins. 
Namely, near the parameter regions 
in which the KN phase is found in Fig.~\ref{PD-variational-calc_all-order}(a), 
we enumerate the energies for $2^{12}$ states 
(including the equivalent states from the symmetry point of view) 
and compare them to find the lowest energy state.
This procedure corresponds to the full parameter search within the 12-sublattice collinear spin configurations. 
We find  that the KN state is the lowest energy state at and around $n=5/8$, while it is not at and around $n=1/8$. 
The result suggests that the (4e) phase near $n=1/8$ 
is taken over by another ordered state with a larger unit cell, but that at and around 5/8 filling 
is stable, at least, up to the 12-site unit cell in the Ising limit.
The stability of the 5/8-filling phase will be further confirmed by the simulated annealing 
in the Heisenberg case in Sec.~\ref{sec:annealing_result}. 
For these reasons, we consider only the 5/8-filling KN state hereafter.

As shown in Fig.~\ref{PD-variational-calc_all-order}(a), 
the (4e) KN state is not limited just at $n=5/8$ but appears in a narrow filling region around 5/8 filling (see also Fig.~\ref{PD_D} in the Appendix~\ref{app:PD_single-ion-anisotropy}). 
This is related to the peculiar electronic structure of this state, as discussed in Sec.~\ref{sec:Dirac}. 
At $n=5/8$, however, the KN phase is stabilized in a wide region of the parameter space of $J_{\rm AF}$ and $J_{\rm H}$. 
Figure~\ref{PD-variational-calc_all-order}(b) shows the ground-state phase diagram at $n=5/8$ as a function of $J_{\rm{AF}}$ and $J_{\rm{H}}$. 
As shown in the figure, the KN phase (4e) widely extends to the large $J_{\rm H}$ region for an intermediate value of $J_{\rm AF}$.

We note that the (4d) phases appear adjacent to the (4e) KN phase, as shown in Fig.~\ref{PD-variational-calc_all-order}(b). 
The (4d) state is a spin-canted version of (4e), as shown in Fig.~\ref{PD-variational-calc_all-order}(c) [(4d) with $\theta=\pi$ corresponds to (4e)]. 
The (4d) phase is interesting because the canted spins give a nonzero spin scalar chirality, leading to the topological Hall effect. 
The possibility of this chiral state was discussed also for the model defined on triangular-kagome lattices~\cite{Akagi2013}.

Let us remark on the validity of the phase diagrams. 
Since the phase diagrams are obtained by the variational calculation while assuming a set of the variational states listed in Fig.~\ref{PD-variational-calc_all-order}(c), 
the results might include artificial phases. 
This should be carefully examined, in particular, in metallic regions where an incommensurate magnetic order with a longer period is anticipated 
[the ferromagnetic (1a) state is rather safe because of the robust DE mechanism].
However, the insulating phases are relatively trustworthy, as they are stabilized by opening an energy gap under the commensurate magnetic order. 
For instance, the (4a) states at $n=1/4$ and $3/4$ were confirmed by several unbiased numerical simulations~\cite{Kumar2010,Kato2010,Barros2013}.
The argument may be applied to the (4e) KN state that we are particularly interested in, because it is a semimetallic state 
with vanishing density of states accompanied by the commensurate magnetic order, as discussed in 
Sec.~\ref{sec:Dirac}. 
Indeed, in addition to the enumeration of all the possible Ising states in the 12-site unit cell discussed above, we will confirm the stability by the simulated annealing in Sec.~\ref{sec:annealing_result}.

\subsection{ \label{sec:Dirac}
Dirac electrons in the kagome network phase
}

\begin{figure}[!htbp]
\begin{center}
\includegraphics[width=0.9\columnwidth]{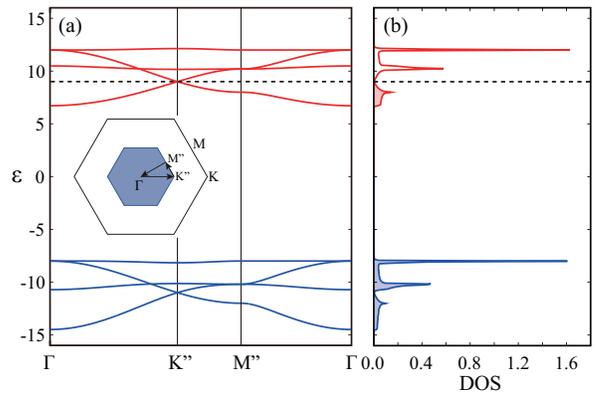}
\end{center}
\caption{(Color online) (a) The band structure and (b) the density of state at 
$J_{\rm{H}}=10$ in the spontaneous KN phase. 
Note that the band structure does not depend on the value of $J_{\rm AF}$
up to a constant energy shift (we take $J_{\rm AF}=0$ here).
The blue (white) hexagon in the inset of (a) indicates the folded (original) Brillouin zone. 
The band dispersions are plotted along the symmetric line 
in the folded Brillouin zone.
The red (blue) color indicates the contributions from the majority (minority) spins.
The dashed line indicates the Fermi level located just at the Dirac node at the ${K''}$ point.
}
\label{band_DOS}
\end{figure}

The KN phase found at $n=5/8$ has a peculiar electronic structure at the Fermi level: The Dirac node with linear dispersion. 
In other words, the system is a Dirac semimetal in the KN phase. 
This is understood as follows. 
First, let us discuss the limit of $J_{\rm H}\to\infty$. 
In this limit, the hopping amplitude between up- and down-spin sites becomes zero: 
The up-spin KN and the isolated down-spin sites are completely separated. 
Consequently, the electronic structure consists of two independent contributions from them. 
The isolated down-spin sites give rise to a flat band at $\varepsilon = \pm J_{\rm H}$. 
Meanwhile, the up-spin KN leads to two copies of the band structure, 
each of which has the same form as that for the non-interacting tight-binding model on the kagome lattice. 
The band structure consists of two dispersive bands 
in the energy range of $-4t\pm J_{\rm H} \le \varepsilon \le 2t\pm J_{\rm H}$, 
in addition to a flat band at $\varepsilon = 2t\pm J_{\rm H}$. 
Each pair of the two dispersive bands 
comprises a Dirac node at $\varepsilon=-t\pm J_{\rm H}$ at the $K''$ point in the Brillouin zone. 
At $n=5/8$, the lower four bands (two dispersive and two flat bands) are fully occupied, while 
the upper bands are occupied only up to the Dirac point; hence, the Fermi level is located at the Dirac node 
in the upper two dispersive bands.

A similar situation is seen in the KN phase for finite but much larger $J_{\rm H}$ than the noninteracting bandwidth.
Figure~\ref{band_DOS} shows the electronic structure in the KN state at $J_{\rm H}=10 $ (note that $J_{\rm AF}$ 
does not change the electronic structure up to a constant energy shift).
As shown in the figure, the electronic structure splits into two sets of bands: 
The upper (lower) set of bands centered at $\varepsilon \simeq J_{\rm H}$ ($\varepsilon \simeq -J_{\rm H}$). 
Each set of bands consists of two parts:
One is similar to the kagome lattice electronic structure with the Dirac node at $\varepsilon \simeq -t\pm 10$, 
and the other is a nearly flat mode at $\varepsilon \simeq \pm 10$, as expected from the above consideration in the limit of $J_{\rm H}\to\infty$. 
Although the form of each band is modified because of 
nonzero hopping between up- and down-spin sites allowed for finite $J_{\rm H}$, the Dirac node with linear dispersion is preserved and the Fermi level is just at the Dirac node, as shown in Fig.~\ref{band_DOS}~\cite{comment1}.

The interesting point of the Dirac node is that the Dirac electrons are almost perfectly spin-polarized, e.g., over $99\%$ of the full moment 
at $J_{\rm H}=10$ (fully polarized in the limit of $J_{\rm H} \to \infty$). 
Namely, this KN state provides a half-semimetal with Dirac electrons.
The Dirac semimetal has been extensively studied from the discovery of graphene, from the potential for applications to electronic devices~\cite{Neto2009}. 
In addition, the half-metallicity will also be beneficial for spintronics, as the electronic state can be manipulated by an external magnetic field. 
In Sec.~\ref{sec:magnetic-field}, we will demonstrate such magnetic field control. 
We note that a similar Dirac half-semimetal was discussed for a three-sublattice ferrimagnetic state at 1/3 filling~\cite{Ishizuka2012}.

Now, let us discuss the stabilization mechanism of the KN phase at $5/8$ filling.
The KN phase is a zero-gap semiconductor 
with the Dirac node at $5/8$ filling, as shown above. 
In many cases, a magnetically ordered phase is stabilized by opening an energy gap. 
Indeed, in the phase diagram in Fig.~\ref{PD-variational-calc_all-order}(a), 
the (4a) scalar chiral phases at 1/4 and 3/4 filling appear with opening an energy gap. 
The Dirac node is not a full gap, but it may give rise to an energy gain for the KN phase. 
The difference between the full-gap insulator and zero-gap semiconductor 
is seen in the phase diagram; 
the gapped chiral phases are limited just at 1/4 and 3/4 filling in the large $J_{\rm H}$ region, as shown in Fig.~\ref{PD-variational-calc_all-order}(a) 
(see also Fig.~3 in Ref.~\cite{Akagi2010}),
whereas the zero-gap semiconducting 
KN phase appears in a very narrow but a finite range of the electron 
filling around $n=5/8$ [see Fig.~\ref{PD-variational-calc_all-order}(a); see also Fig.~\ref{fig:app} in the Appendix~\ref{app:PD_single-ion-anisotropy}]. 
The narrow but nonzero width in $n$ 
might be a footprint of the fact that 
the Dirac node formation contributes to the stabilization of the KN phase.

\subsection{ \label{sec:annealing_result}
Stability of the kagome network phase: Simulated annealing results
}

Next, we examine the stability of the KN phase at 5/8 filling within the parameter space where the phase emerges 
in Fig.~\ref{PD-variational-calc_all-order}(b)
by the simulated annealing for the enlarged 12-sites magnetic unit cell, as described in Sec.~\ref{sec:simulated_annealing}.
We find that it is difficult to obtain a converged result at 5/8 filling, 
suggesting (quasi)degeneracy with other states or the possibility of a larger sublattice order than the 12 sites~\cite{comment2}. 
In Sec.~\ref{sec:vatiational_calculation}, however, we found that the KN state is the lowest-energy state in the Ising limit within the 12-site unit cell. 
We, therefore, performed the simulated annealing by adding the spin anisotropy in 
the model~(\ref{Ham:DE}) as 
\begin{eqnarray}
{\cal H}={\cal H}_{\rm KL}-D 
\sum_{i} (S^z_i)^2,
\label{Ham:DE_D}
\end{eqnarray}
where the first term is the Hamiltonian in Eq.~(\ref{Ham:DE}) and the second term 
denotes the single-ion anisotropy ($D > 0$); $S_i^z$ is the $z$ component of $\mathbf{S}_i$. 
Note that the limit of $D \to \infty$ corresponds to the Ising limit.
In the variational calculations, we confirmed that $D$ further stabilizes the KN state; see the Appendix~\ref{app:PD_single-ion-anisotropy} for details.

\begin{figure}[!htbp]
\begin{center}
\includegraphics[width=0.97\columnwidth]{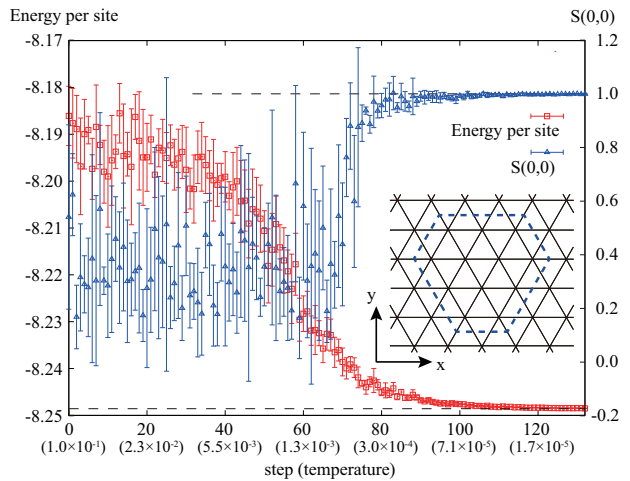}
\end{center}
\caption{(Color online) Energy per site 
(left axis) and the spin structure factor 
$S(\mathbf{q})$ at $\mathbf{q}=(0,0) $ (right axis) as functions of the cooling step in the simulated annealing 
for $J_{\rm{H}}=10$,  $J_{\rm{AF}}=0.12$, and $D=0.044$ at $n=5/8$.
The dashed lines indicate the values of energy and $S(0,0)$ in the limit of $T=0$ in the KN state.
See the text for details.
In the inset, the blue dashed hexagon on a triangular lattice represents the 12-site unit cell on which we performed the simulated annealing.
}
\label{SA_SI}
\end{figure}

We find that, by introducing a small single-ion anisotropy $D$, the simulated annealing gives a converged solution to the KN state. 
As for the demonstration, in Fig.~\ref{SA_SI}, we show the cooling process in the simulated annealing at $D=0.044$. 
The calculation was done at $J_{\rm{H}}=10$, $J_{\rm{AF}}=0.12$, and $n=5/8$. 
Figure~\ref{SA_SI} shows the energy per site, $\langle {\cal H} \rangle/N$, and the spin structure factor $S(\mathbf{q})$ at 
$\mathbf{q}=(0,0) $ as functions of the cooling step (temperature) in the simulated annealing. 
The spin structure factor is defined as $S(\mathbf{q})=\frac{1}{N} \sum_{i,j} \mathbf{S}_{i} \cdot \mathbf{S}_{j} \exp{(i\mathbf{q}\cdot \boldsymbol{r}_{ij})}$, where $\boldsymbol{r}_{ij}$ denotes the position vector from $i$th to $j$th site.
The KN state is signaled by peaks of $S(\mathbf{q})$ at $\mathbf{q}=(0,0)$, $\pm (0,2\pi/\sqrt{3})$, $\pm (\pi,-\pi/\sqrt{3})$, and $\pm (\pi,\pi/\sqrt{3})$ with equal weights.
As shown in the figure, the energy and $S(0,0)$ converge 
to the values expected in the KN state (dashed lines) as $T \to 0$.
During the cooling process, $S(0,2\pi/\sqrt{3})$, $S(\pi,-\pi/\sqrt{3})$, and $S(\pi,\pi/\sqrt{3})$ have the same values as $S(0,0)$ within the error bars. 
We performed similar runs for a total of $10$ different parameters sets and always found the same behavior as in Fig.~\ref{SA_SI}.
The results clearly indicate that the KN state found at 5/8 filling in the variational calculation remains stable within the 12-site unit cell, at least, in the presence of a 
small Ising-type anisotropy.

\subsection{ \label{sec:mean-field_result}
Effect of off-site Coulomb repulsion
}
As discussed in Sec.~\ref{sec:Dirac}, in the KN phase at 5/8 filling in the large $J_{\rm H}$ region, 
the lower four bands are fully occupied and 
the upper bands are occupied up to the Dirac node. 
This electronic state is approximately regarded as the 1/3-filling state in the spinless fermion model on a kagome lattice: 
The doped electrons in the upper bands are almost confined in the 
KN with aligning their spins antiparallel to the localized up-spins because of the large $J_{\rm H}$, and the band filling corresponds to 1/3 
in the bands originating from the KN. 
The mapping becomes exact in the limit of $J_{\rm H}\to \infty$. 
In the spinless fermion model at 1/3 filling,
it was pointed out that a fractional charge excitation ($\pm e/2$; $e$ is the elementary charge) emerges when the system has a 
strong Coulomb repulsion between nearest-neighbor sites~\cite{OBrien2010}. 

Bearing such a possibility in mind, we investigate the effect of
a nearest-neighbor Coulomb repulsion on the KN formation. 
The Hamiltonian is given by
\begin{eqnarray}
{\cal H}={\cal H}_{\rm KL}+V \! \sum_{\langle i,j \rangle} \! n_i n_j,
\end{eqnarray}
where $V$ is the Coulomb repulsion between nearest-neighbor sites, and $n_i=\sum_{\alpha} c^{\dagger}_{i,\alpha } c_{i,\alpha }$ is
a number operator for itinerant electrons on site \textit{i}.
Here, we study the effect of the $V$ term by means of the mean-field approximation. 
Namely, we decouple the interaction term by using the standard Hartree-Fock approximation as 
\begin{align}
  n_{i}n_{j} & \approx  n_{i}\langle n_{j}\rangle + \langle n_{i}\rangle n_{j} - \langle n_{i}\rangle \langle n_{j}\rangle  \notag\\  
  &- \sum_{\alpha\beta} \left( c_{i\alpha}^{\dagger}c_{j\beta}\langle c_{j\beta}^{\dagger}c_{i\alpha}\rangle + \langle c_{i\alpha}^{\dagger}c_{j\beta}\rangle c_{j\beta}^{\dagger}c_{i\alpha} \right.  \notag\\ 
  &- \left. \langle c_{i\alpha}^{\dagger}c_{j\beta}\rangle \langle c_{j\beta}^{\dagger}c_{i\alpha}\rangle \right). \label{eq:decouple}
\end{align}
In the mean-field analysis, 
we assumed that configurations of the classical localized spins are (1a), (2a), (3a), (3d), and (4e) in Fig.~\ref{PD-variational-calc_all-order}(c), as these orderings appear in Fig.~\ref{PD-variational-calc_all-order}(b).
In the calculation, we take $200 \times 200$ grid points in the folded first Brillouin zone.

\begin{figure}[!htbp]
\begin{center}
\includegraphics[width=0.95\columnwidth]{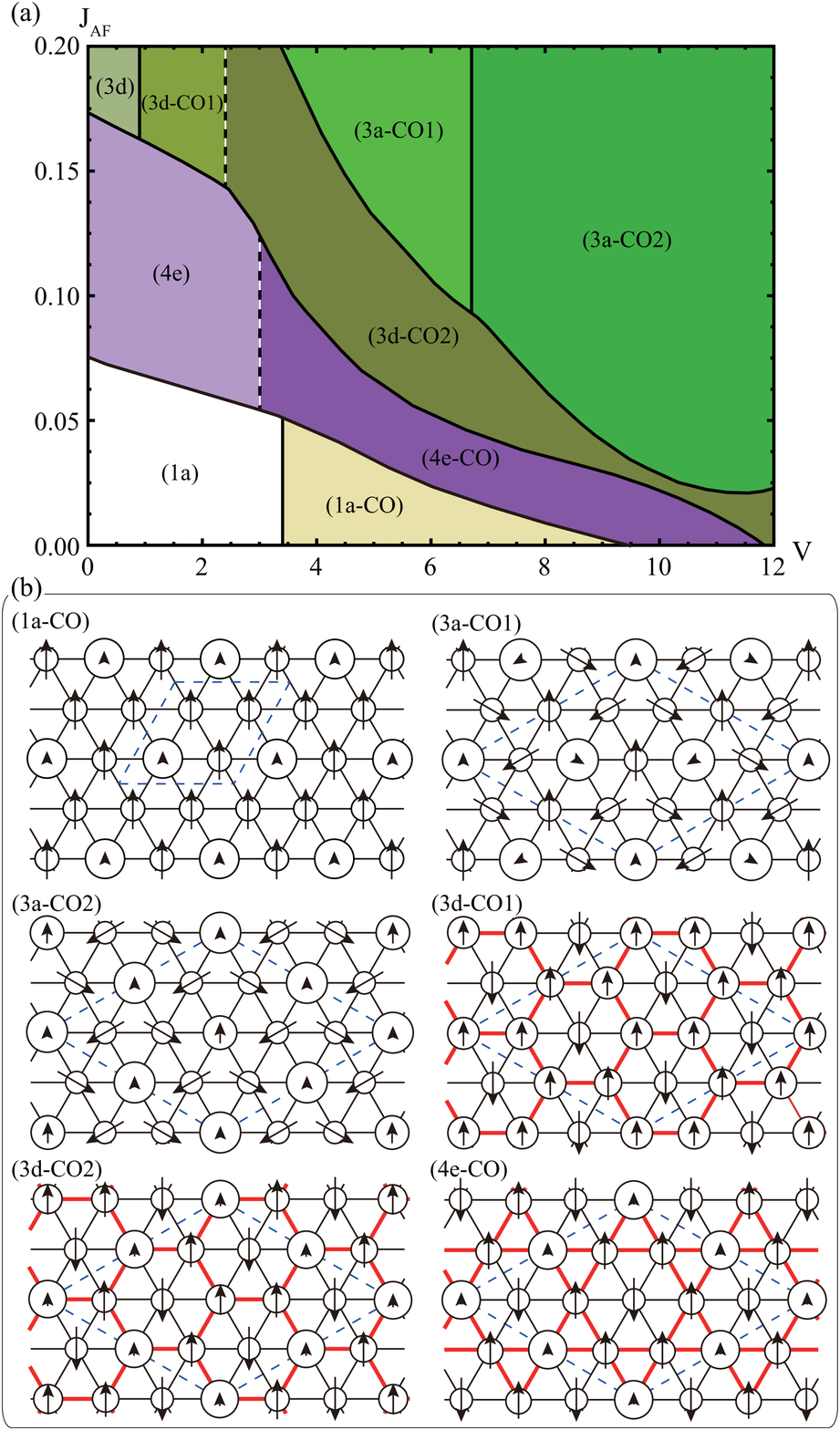}
\end{center}
\caption{(Color online) (a) Ground-state phase diagram obtained by the mean-field approximation as a function 
of $V$ and $J_{\rm{AF}}$. 
The calculation was done for $J_{\rm{H}}=20$ at $n=5/8$.
The notations such as (1a) and (3d) are common to those in Fig.~\ref{PD-variational-calc_all-order}, which label the magnetic ordering patterns. 
(3d) ordered region of the phase diagram is a collinear order with $\theta =\pi$ except at $V=0$.
The vertical dashed lines indicate the second-order transition.
(b) Schematic pictures of the order patterns 
of the phases in (a):
(1a-CO) four-sublattice FM $\mathbf{q}=\mathbf{0}$ charge order (CO), 
(3a-CO1) 12-sublattice $120^{\circ }$-AFM $\mathbf{q}=\mathbf{0}$ CO, 
(3a-CO2) 12-sublattice $120^{\circ }$-AFM CO with a nonzero net moment, 
(3d-CO1) and (3d-CO2) 12-sublattice ferrimagnetic collinear CO with nonzero net moments, 
and (4e-CO) 12-sublattice KN $\mathbf{q}=(2\pi/3,2\pi/\sqrt{3})$ CO with a nonzero net moment. 
The size of circles (length and direction of arrows) denotes the charge density 
(the length and direction of spins) of itinerant electrons.
The blue dashed rhombuses represent the unit cells. The red lines are the up-spin networks.
All the directions of localized spins are parallel to those of itinerant electrons spins.
All the charge ordered phases are insulating. 
}
\label{PD_V}
\end{figure}

Figure~\ref{PD_V} shows the ground-state phase diagram at 5/8 filling obtained by the mean-field approximation. 
At $V=0$, there are three phases, (1a), (4e), and (3d), while changing $J_{\rm AF}$. 
The (4e) and (3d) states exhibit a charge disproportionation even at $V=0$, because of the translational symmetry breaking by magnetic ordering; 
the local charge density at up-spin sites are higher than that at down-spin sites. 
While increasing $V$, however, an additional translational symmetry breaking appears in each region by charge ordering, i.e., a superstructure of charge density distinct from the magnetic one. 
From the arguments in the last part in Sec.~\ref{sec:vatiational_calculation}, 
we regard that, at least, the (4e-CO) and (1a-CO) phases are trustworthy as they evolve from 
the reliable (4e) and (1a) states at $V=0$, respectively.

Among various charge ordered phases, 
we are interested in the (4e-CO) phase, which emerges from the (4e) KN phase through a second-order phase transition while increasing $V$. 
The (4e-CO) phase is the KN state accompanied by charge ordering with the wave number 
$\mathbf{q}=(2\pi/3,2\pi/\sqrt{3})$, that is the so-called $\sqrt{3} \times \sqrt{3}$ type. 
The charge ordering pattern is the same as that discussed for the spinless fermion model 
on a kagome lattice at $1/3$ filling~\cite{Nishimoto2010}.
The charge-ordered state in the spinless fermion system can exhibit fractional charge excitations
when an electron (a hole) is added to the 1/3-filling state~\cite{OBrien2010}. 
Therefore, it is expected that our mean-field state (4e-CO) is connected to the phase with fractional charge excitations. 
In fact, a ``defect" by electron (hole) doping in the (4e-CO) state can propagate without energy loss, 
at least for $V \gg t$ (see also Fig.~4 in Ref.~\cite{OBrien2010}). 
Although it is necessary to go beyond the mean-field approximation to clarify 
whether the KN state accommodates fractional charge excitations, 
this is an interesting issue since fractional charge excitations usually emerge 
in the systems whose lattice structures consist of corner-sharing units. 
The possibility in the current system on the edge-sharing triangular lattice is left for future study.

\subsection{ \label{sec:magnetic-field}
Effect of external magnetic field
}
\begin{figure}[!htbp]
\begin{center}
\includegraphics[width=0.9\columnwidth]{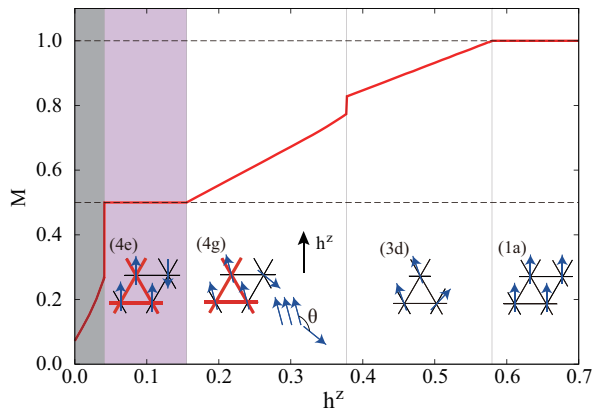}
\end{center}
\caption{(Color online) 
Magnetization per site as a function of an external magnetic field $h^z$ in the $z$ direction 
obtained by the simulated annealing supplemented by variational calculations. 
The data are calculated for $J_{\rm H}=10$ and $J_{\rm AF}=0.12$ at $n=5/8$.
The (4e), (4g), (3d), and (1a) phases appear for $0.04 \lesssim h^z \lesssim 0.16$, $0.16 \lesssim h^z \lesssim 0.38$, $0.38 \lesssim h^z \lesssim 0.58$, and $0.58 \lesssim h^z $, respectively.
The first-order transition occurs near $h^z \sim0.38$. Others are continuous transition.
In the shaded area for $0 \le h^z \lesssim 0.04$, 
it is hard to obtain good convergence; 
we plot the magnetization for the 12 sublattice chiral order~\cite{comment2} appearing as the solution in the simulated annealing. 
}
\label{magnetization-curve}
\end{figure}

Finally, we discuss the effect of an external magnetic field on the KN state.
We consider the Zeeman coupling to a magnetic field applied in the $z$ direction,
\begin{eqnarray}
{\cal H}_{\rm Z}=-h^z \! \sum_{i} S^z_i,
\label{eq:H_hz}
\end{eqnarray}
as an additional term to Eq.~(\ref{Ham:DE}).
For simplicity, we neglect the coupling of orbital motion of itinerant electrons to the magnetic field
as well as the Zeeman coupling to itinerant electrons. 

We investigate the effect of the Zeeman term in Eq.~(\ref{eq:H_hz}) on the model in Eq.~(\ref{Ham:DE}) by using the simulated annealing. 
We find that the external magnetic field stabilizes the KN phase 
as in the case of the single-ion anisotropy $D$ discussed in Sec.~\ref{sec:annealing_result}.
Figure~\ref{magnetization-curve} shows the magnetization curve calculated by 
combining the simulated annealing and variational calculations~\cite{comment3}. 
Here, the magnetization is defined by 
$M=\sum_{i} S^z_i / N$. 
As shown in the figure, the KN phase appears for $0.04 \lesssim h^z \lesssim 0.16$, with exhibiting magnetization plateau at $M=1/2$. 
Further increasing $h^z$, the KN phase changes into the 
(4g) phase which is a canted 
version of the KN state for $0.16\lesssim h^z\lesssim0.38$,
(3d) phase for $0.38\lesssim h^z\lesssim0.58$~\cite{comment4}, 
and finally turns into the FM phase for $h^z\gtrsim0.58$ ($\sim5J_{\rm AF}$). 
The transitions at $h^z \sim 0.16$ and $0.58$ are both continuous, whereas that at $h^z \sim 0.38$ is of first order. 
We note that both the (4g) 3:1 and (3d) 2:1 canted phases are magnetic counterparts of supersolid~\cite{Liu1973, Ueda2010}. 

The result suggests an interesting controllability of the Dirac electronic state, 
i.e., one can destroy and create the Dirac electrons by sweeping the magnetic field.
While increasing the magnetic field, it shows a transition to a magnetic supersolid phase, in which  exotic behavior is expected as well.
Furthermore, once the KN phase exhibits fractional charge excitations in the presence of $V$ (see Sec.~\ref{sec:mean-field_result}), 
there is the possibility of controlling the appearance of fractional charge excitations by magnetic field.

\section{ \label{sec:summary}
Summary
}
We have investigated the Kondo lattice model on a triangular lattice by using the variational calculation and the simulated annealing method.
We found that the system exhibits the kagome network state 
at a special filling $n=5/8$ for the large Hund's-rule coupling.
In the kagome network state, the system is spontaneously divided into two parts by the four-sublattice collinear ferrimagnetic order: One is the kagome lattice composed of up-spin sites, and the other is isolated down-spin sites in the hexagons of the kagome lattice.
This peculiar magnetism gives rise to an unusual electronic state: A Dirac half-semimetallic state.
The semimetallic dip of the density of states due to the Dirac node formation contributes to the stabilization of the kagome network phase.
We have also found that 
a $\sqrt{3}\times\sqrt{3}$-type charge order occurs in the 
kagome network state as the mean-field solution when 
we include the Coulomb repulsion between nearest-neighbor sites. 
This charge ordering pattern is the same as that 
for the spinless fermion model on a kagome lattice at $1/3$ filling discussed 
from the interest of fractional charge excitations.
Moreover, we have found that the emergence of the kagome network Dirac 
half-semimetal can be controlled by an external magnetic field.
Considering real materials, the parameters used in the present calculations are reasonable to some extent:
The Hund's-rule coupling $J_{\rm H}$ is dozens of times larger than the transfer integral, 
and it is one or two orders of magnitude larger than the antiferromagnetic superexchange interaction $J_{\rm AF}$ and the single-ion anisotropy $D$.
These energy scales are seen in some real materials, such as manganese oxides. 
It is desired to explore the kagome network phase discussed in the paper 
in real systems as a candidate for the peculiar Dirac half-semimetal and possible fractional charge excitations.

\acknowledgements
{
We acknowledge helpful discussions with Chisa Hotta, Takahiro Misawa, and Masafumi Udagawa.
Y.A. thanks Satoru Hayami, Ryui Kaneko, Yuichi Motoyama, 
Junki Yoshitake, Hiroaki T. Ueda, and Nic Shannon for helpful comments. 
Y.A. was supported by Grant-in-Aid for JSPS Fellows. 
Y.A. acknowledges support from Okinawa Institute of Science and Technology Graduate University. 
This work was supported by Grants-in-Aid for Scientific Research (Grant Nos. 21340090, 22540372, and 24340076)
Global COE Program ``the Physical Sciences Frontier,"
the Strategic Programs for Innovative Research (SPIRE),
MEXT, and the Computational Materials Science Initiative (CMSI), Japan.
}

\appendix

\section{Variational Phase Diagram in the Presence of the Single-ion Anisotropy}
\label{app:PD_single-ion-anisotropy}

\begin{figure}[H]
\begin{center}
\includegraphics[width=0.85\columnwidth]{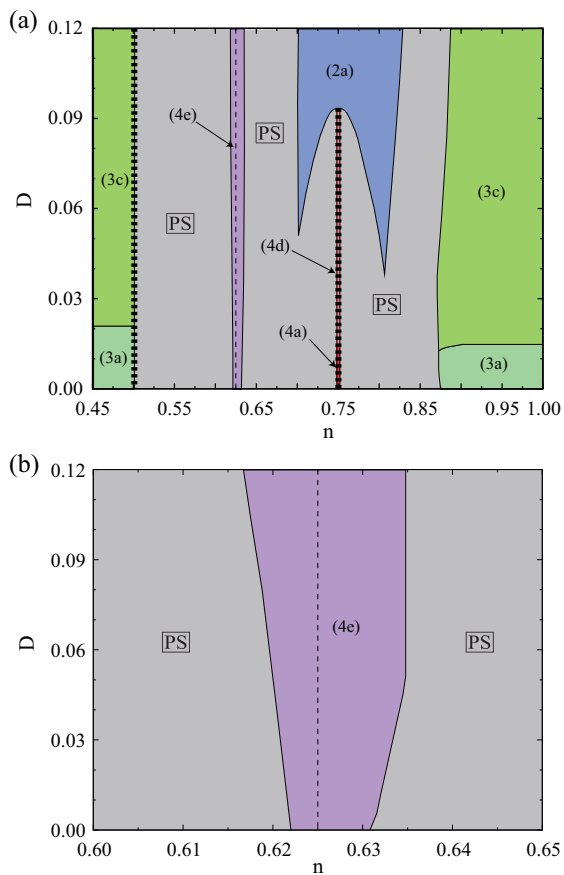}
\end{center}
\caption{(Color online) 
Ground-state phase diagrams for the model (\ref{Ham:DE_D}) 
as functions of $n$ and $D$ at $J_{\rm H}=10$ and $J_{\rm{AF}}=0.12$ obtained by the variational calculation.
(b) is the enlarged view of (a) near $n=5/8$.
The vertical thick (thin) dashed lines at $n=1/2$ and $3/4$ ($n=5/8$) in (a) indicate gapful insulating 
(zero-gap semiconducting) regions. 
PS stands for the phase-separated region.
\label{fig:app}
}
\label{PD_D}
\end{figure}

In this Appendix, to clarify the effect of the single-ion anisotropy, we study the ground-state phase diagram for the model in Eq.~(\ref{Ham:DE_D}) by the variational calculations. 
Using the method in Sec.~\ref{sec:variational_calculation_method}, we performed the calculations around $n=5/8$, as we are interested in the stability of the KN phase at $5/8$ filling.

Figure~\ref{PD_D}(a) shows the result of the ground-state phase diagram as a function of the electron filling $n$ and the single-ion anisotropy $D$ at $J_{\rm H}=10$ and $J_{\rm AF} = 0.12$.
We find that the (4e) KN phase remains stable around $n=5/8$ in the presence of $D$~\cite{comment5}. 
Indeed, it is further stabilized by increasing $D$: The width of the KN phase in $n$ becomes wider for larger $D$, as clearly shown in the enlarged figure in Fig.~\ref{PD_D}(b). 
Although it is reasonable that the single-ion anisotropy stabilizes the collinear magnetic order, the result is not trivial because the width is determined by the grand potential for the competing phases on the other sides of the 
PS regions. 
In the present case, the competing phases are the (3a) phase at $n=1/2$ and the (4a) phase at $n=3/4$ in the small $D$ region.
Increasing $D$, the (3a) and (4a) phases 
change into their spin canting versions, 
namely, (3c) and (4d), 
respectively, as shown in Fig.~\ref{PD_D}(a).
With further increasing $D$, the (4d) phase 
is eventually taken over by the metallic (2a) phase with a two-sublattice collinear order.
Our results indicate that the region of the (4e) KN phase is extended as $D$ increases despite the competition to these phases.


\begin{thebibliography}{60}
\bibitem{comment1} We note that the (4d) phase near $J_{\rm AF} = 0.16$ in Fig.~\ref{PD-variational-calc_all-order}(b) also shows the Dirac node at 5/8 filling. On
the other hand, the (4d) phase near $J_{\rm AF} = 0.08$ does not have a Dirac node in the electronic structure. 
This is because the canting angle is small ($\sim 30^{\circ }$) for the latter, while it is large ($\sim 160^{\circ }$) for the former.
\bibitem{comment2} A candidate for the lowest-energy state found in the simulated annealing is
a 12-sublattice state with a noncoplanar spin configuration.
This state is composed of four triangles of ferromagnetically aligned three spins;
the four moments comprise an all-out-type configuration.
\bibitem{comment3} 
The overall magnetization curve is obtained by the simulated annealing. 
The result, however, includes small numerical noises, which are presumably due to the small $xy$ spin components remaining in the converged solutions. 
We therefore finalize the magnetization curve by using the variational states by assuming that the net $xy$ component vanishes.
\bibitem{comment4} We note that for $0.38\lesssim h^z\lesssim0.58$ the (3d) state is almost degenerate in energy with (3c). The (3c) phase is also a magnetic supersolid similarly to (3d).
\bibitem{comment5} In this variational phase diagram, the KN phase remains stable down to $D=0$ [see also Fig.~\ref{PD-variational-calc_all-order}(a)]. We note, however, that in the simulated annealing, the small anisotropy $D$ is required to stabilize the KN phase, as described in Sec.~\ref{sec:annealing_result}.
\end{thebibliography}
\end{document}